\documentclass[epj]{svjour}
\usepackage{graphicx}
\usepackage{multirow}
\usepackage{amsmath,amsfonts,amssymb}
\usepackage{hyperref}
\usepackage{xcolor}
\begin{document}

\title{On the Positronium g-factor}
%\author{J. Agil \and R. Battesti \and C. Rizzo% etc

\author{J. Agil$^1$ \and D. Bakalov$^2$ \and  R. Battesti$^1$ \and C. Rizzo$^1$ %etc.
}

\author{J. Agil\inst{1}\inst{*} \and R. Battesti\inst{2} \and C. Rizzo\inst{2} \and D. Bakalov\inst{3}}
\institute{ 
\inst{1} CNRS, LNCMI UPR 3228 (UGA, UT3, INSA-T, EMFL), F-31400 Toulouse Cedex, France\\
\inst{2} Université Toulouse 3, LNCMI UPR CNRS 3228
(UGA, INSA-T, EMFL), F-31400 Toulouse Cedex, France\\
\inst{3} Institute for Nuclear Research and Nuclear Energy, Bulgarian Academy of Sciences, blvd. Tsarigradsko ch. 72, Sofia 1142, Bulgaria \\
\inst{*} Corresponding author: jonathan.agil@lncmi.cnrs.fr
}

%\thanks{e1}{e-mail: fauthor@example.com}

%\offprints{J. Agil}          % Insert a name or remove this line
%
%\institute{Laboratoire National des Champs Magn\'{e}tiques Intenses (UPR 3228, CNRS-UPS-UGA-INSA), 143 avenue de Rangueil, Toulouse, 31400, France 
%\and
%Institute for Nuclear Research and Nuclear Energy, Bulgarian Academy of Sciences, blvd. Tsarigradsko ch. 72, Sofia 1142, Bulgaria 
%}

%\institute{Laboratoire National des Champs Magn\'{e}tiques Intenses (UPR
%3228, CNRS-UPS-UGA-INSA), 143 avenue de Rangueil,
%Toulouse, 31400, France.}

%
\date{Received: date / Revised version: date}

\abstract{In this letter, we recall the main facts concerning the g-factor of positronium and we show how the value of the g-factor of the positronium is important. Taking it better into consideration may provide a solution to the reported discrepancy between QED theory and experiments concerning the hyperfine splitting of the fundamental level of the positronium.  We also give the only experimental value that existing experiments can provide, $g_{\mathrm{Ps}}=2.0023\pm 0.0012$ at $3\sigma$.  
 % end of PACS codes
} %end of abstract
\maketitle
Positronium (Ps) is the name usually given to the bound state of an electron and its antiparticle, the positron. It is a special case of hydrogen-like atom with the particularity that the ratio of the mass of the two components is exactly equal to one and that the only interactions between the two particles are electromagnetic or weak. It is therefore a pure QED system and a very important test-bench for QED theory and the related experiments~\cite{Cassidy2018,Adkins2022}. 

Typically these experiments consist in spectroscopy measurements, often performed in the presence of a magnetic field~\cite{Arimondo2016} as in the case of the positronium HyperFine Splitting (HFS) measurements. 
In this particular case, discrepancies between theory prediction and experimental results has been reported both for the $n=1$~\cite{Arimondo2016} and $n=2$~\cite{Gurung2020} levels. 
%<<<<<<<<<<<<<<<<<<

In the rest frame, in the lowest order, the interaction of a general hydrogen-like system (atom or ion) with an 
external magnetic field $B$  
%in lowest order 
is described with the Zeeman Hamiltonian $H_Z$, which may be put in the form
\begin{equation}\label{Zeeman}
H_Z = \mu_B(g_L \mathbf{L} + g_S \mathbf{S} + g_I^\prime \mathbf{I})\cdot \mathbf{B} = \mu_B\,g_J \,\mathbf{J} \cdot\mathbf{B},
\end{equation}
where $\mathbf{S}$ and $\mathbf{I}$ are the electron and nuclear spin angular momenta, 
$\mathbf{L}$ is the atom orbital momentum,  and $\mathbf{J}$ is the total angular momentum.
In formula~\ref{Zeeman} we have introduced the dimensionless g-factors: $g_S$ for the electron spin, $g_I^\prime$ for the nuclear spin, $g_L$ for the atomic orbital momentum, and $g_J$ for the total angular momentum, all of which are measured in units of the Bohr magneton, $\mu_B$. In this notation, $g_L$ and $g_S$ are assumed to be positive, and $g_I^\prime$ negative~\cite{Arimondo2016}. The numerical values of all g-factors are system- and state-dependent.

The bound electron g-factor $g_S$ is different from $g_e$, the g-factor of free electron that is known very precisely, $g_e = 2.00231930436256(35)$~\cite{CODATA}. For example, for an electron bound to a point like nucleus with charge $Z$ by a pure Coulomb potential the g-factor for the fundamental state is at the lowest order~\cite{Breit1928}
\begin{equation}\label{gposBreit}
g_{S} = g_e\left(1-\frac{1}{3}(Z\alpha)^2\right).
\end{equation}
In the case of the positronium, $g_I^\prime = - g_S$. Moreover, in the ground state $L=0$, the Zeeman Hamiltonian Eq.~\ref{Zeeman} takes the form  
\begin{equation}
H_Z=\mu_B\,g_{\mathrm{Ps}}\,(\mathbf{S}-\mathbf{I})\cdot\mathbf{B},
\label{Zeeman-Ps}
\end{equation}
where $g_{\mathrm{Ps}}=g_S$ denotes the only g-factor relevant in this case.
The value of $g_{\mathrm{Ps}}$ has been calculated by Grotch and Kashuba~\cite{Grotch1973} and Lewis and Hughes~\cite{Lewis1973} both in 1973 and more recently by Anthony and Sebastian in 1994~\cite{Anthony1994}. 
%It is only given for the $n^1S_{1/2}$ or $2^3P_j$ states at $\alpha^2$ level. 
For the fundamental level $1S_{1/2}$~\cite{Grotch1973} and~\cite{Anthony1994} give:
\begin{equation}\label{gposSeb}
\begin{split}
g_{\mathrm{Ps}} &= 2\left(1+a_e -\frac{5}{24}\alpha^2 -\frac{\alpha^2}{24}a_e\right)\\
&= g_e\left(1 -\frac{5}{24}\alpha^2 +\frac{1}{6}\alpha^2 a_e\right)+o(\alpha^3),
\end{split}
\end{equation}
where  $a_e = (g_e-2)/2$.
Following~\cite{Grotch1973}, neglected terms should be of the order $\alpha^4$  \emph{i.e.} $g_{\mathrm{Ps}}$ theoretical prediction should be known at about $3\times 10^{-9}$. 

A slightly different formula is given in~\cite{Lewis1973} 
\begin{equation}\label{gposLew}
g_{\mathrm{Ps}} = g_e\left(1-\frac{5}{24}\alpha^2 -\frac{T}{2mc^2}\right),
\end{equation}
where $T$ is the total kinetic energy of the atom and $mc^2$ the electron rest energy. %-------------------

The considerations in \cite{Lewis1973} are based on a relativistic Hamiltonian of positronium, which includes the leading relativistic effects of the motion of the center-of-masses of the system.
In particular, the value of $g_{\rm Ps}$ of Eq.~\ref{gposLew} is expressed in terms of the mean value in the ground state of positronium of the terms in the Hamiltonian that are linear in $\mathbf{B}$ (therefore, the $T$-dependence of $g_{\rm Ps}$).
In the general case the Hamiltonian of \cite{Lewis1973} includes terms that describe the Zeeman and the motional Stark effects, but in the ground positronium state the latter give zero contribution.
The kinetic energy does not appear in Eq.~\ref{gposSeb} because at a rather early stage the derivations in \cite{Anthony1994} were restricted to the positronium center-of-mass rest frame. In the meantime expression~\ref{gposSeb} includes a QED correction of order $O(\alpha^2a_e/6)\sim 10^{-8}$ that was not considered in~\cite{Lewis1973}.

%-------------------
Expressions~\ref{gposSeb} and~\ref{gposLew} are complementary, not alternative; the above two corrections are additive and should both be included in the expression that claims to be accurate up to terms of order $10^{-8}$
\begin{equation}\label{gposCombined}
g_{\mathrm{Ps}} = g_e\left(1-\frac{5}{24}\alpha^2+\frac{1}{6}\alpha^2a_e -\frac{T}{2mc^2}\right).
\end{equation}
%-------------------

Note that the ``kinetic energy'' correction  (of the order of $3\times 10^{-8}$ for thermalized positronium) has some similarity with the Doppler shift, \textit{e.g.} the statistically distributed values of $T$ broaden the positronium energy levels and 
$\Delta\nu$. 
The physical nature of these two effects, however, is completely different.
In the present work terms involving the kinetic energy are only used to estimate the uncertainty of $g_{\rm Ps}$ and $\Delta\nu$.
%-------------------

As a matter of fact, the g-factor of positronium is a fundamental constant whose value has been predicted precisely for the fundamental level. The test of such a value corresponds therefore to an important verification of our understanding of the phenomena related to particle annihilation. So, what is its experimental value? None because $g_{\mathrm{Ps}}$ has never been measured.

The reason is that $g_{\mathrm{Ps}}$ plays an important role in the measurement of the hyperfine splitting (HFS) in the fundamental level of positronium $\Delta\nu$. 
%<<<<<<<<<<<<<<<<<<<<<<<<<<<<<<<
% Insertion
In the leading order in $\alpha$ the hyperfine structure of Ps is described  by the 
spin-dependent terms of the Breit-Pauli Hamiltonian $H_{\rm BP}$ and the annihilation term $H_{A}$ 
\begin{align}
&H_{\rm Ps}=H_{\rm BP}+H_{A},\ \ H_{\rm BP}= H_{F}+H_{so}+H_{t},
\end{align}
where $H_{F}$, $H_{so}$, and $H_{t}$ are respectively the Fermi, 
spin-orbit and tensor spin interactions between the electron and the positron. 
The hyperfine states of Ps, $\vert nLFJJ_z\rangle$,
are labelled with the quantum numbers of the total spin $\mathbf{F}=\mathbf{S}+\mathbf{I}$, of the orbital momentum $\mathbf{L}$, the total angular momentum $\mathbf{J}$ and its projection on the quantization axis $m_J$. 

Using these notations the annihilation term is taken in the form \cite{Lewis1973}
$H_{A}=(\pi/2)\left(e\hbar g_e/mc\right)^2
(\mathbf{S}\cdot\mathbf{I}+3/4) \delta(\mathbf{r})$.
For $s$-states $L=0$, $m_J\!=\!m_F$, and $J=F$ may be omitted; in the ground state $n\!=\!1, L\!=\!0$ and $\vert nLFJm_J\rangle\equiv\vert Fm_F\rangle$. 

In external magnetic field the Hamiltonian of positronium $H_{\rm Ps}$ must include both the spin and magnetic interactions:
\begin{equation}
H_{\rm Ps}=H_{\rm BP}+H_A+H_Z,
\label{hfs+Z}
\end{equation} 
and the hyperfine energy levels $E^{\rm hf}_{nLFJm_J}$ are calculated in first order of perturbation theory by diagonalizing the matrix of $H_{\rm Ps}$.
%>>>>>>>>>>>>>>>>>>>>>>>>>>>>>>>
%
To measure $\Delta\nu=(E^{\rm hf}_{10110}-E^{\rm hf}_{10000})/h$ experimentalists, since~\cite{Deutsch1952}, measure typically the hyperfine transition frequency $f_0=(E^{\rm hf}_{10111}-E^{\rm hf}_{10110})/h$ between $\vert F=1, m_F=\pm1 \rangle$ level, which is not affected by the magnetic field, and $\vert F=1, m_F=0 \rangle$. 
In this case, $f_0$ can be written using the formula~\cite{Deutsch1952,Breit1931,Anthony1994},
\begin{equation}\label{Breit}
f_0 = \dfrac{\Delta\nu}{2}\left[ \sqrt{1+\left(\dfrac{2g_{\mathrm{Ps}}\mu_BB}{h\Delta\nu}\right)^2}-1\right].
\end{equation}
One clearly sees that $f_0$ depends on two fundamental parameters: $\Delta\nu$ and $g_{\mathrm{Ps}}$. Both can be calculated in the framework of QED. As for $g_{\mathrm{Ps}}$, we have discussed it previously, as for $\Delta\nu$, the theoretical prediction is $\Delta\nu_{th} = 203.39169(41)$~GHz~\cite{Adkins2015} i.e. $\Delta\nu$ theoretical prediction is known at about 2 ppm. 

In table~\ref{tab:exp} we resume the experimental data reported by the most recent experiments whose goal was to measure $\Delta\nu$~\cite{Mills1975,Egan1977,Carlson1977,Ritter1984,Ishida2014,Miyazaki2015}. For each of them, we list $f_0$, $B$, the $\Delta\nu$ given and its relative error $\delta\Delta\nu/\nu$. Let's note that the values in the table for $\Delta\nu$ concerning~\cite{Mills1975,Egan1977} are the ones re-evaluated by Mills in 1983~\cite{Mills1983} and that the value of $\Delta\nu$ reported in~\cite{Miyazaki2015} corresponds to a direct measurement of the hyperfine splitting at $B=0$~T.

These values are self-consistent \emph{i.e.} if one introduces the values of the experimental parameters $B$ and $\Delta\nu$ given in the table in the equation~\ref{Breit}, one obtains the $f_0$ value also given in the table. Let us note also that, as far as the magnetic field is concerned, the last digit given in the table corresponds to the precision of the shielded proton gyromagnetic factor at the time of the experiment~\cite{CODATA73,CODATA10}.

\begin{table}
\centering
\resizebox{\linewidth}{!}{
\begin{tabular}{|c|c|c|c|c|}
 \hline
  \rule{0pt}{2.5ex} Experiment & $f_0$ (GHz) & $B$ (T) & $\Delta\nu$ (GHz) & $\delta\Delta\nu/\nu$ (ppm) \\

 \hline
 %\rule{0pt}{2.5ex} 1975~\cite{Mills1975} & $3.25276$  & $2.32$ & $203.3870(16)$ & 8\\
 \hline
 %\rule{0pt}{2.5ex} 1977~\cite{Egan1977} & 2.3 & 0.7761132 & 203.3849(12) & 6\\
 %\hline
 \rule{0pt}{2.5ex} 1977~\cite{Carlson1977} & 2.32 & 0.779516 & 203.384(4)& 20 \\
 \hline
 \rule{0pt}{2.5ex} 1983~\cite{Egan1977}\cite{Mills1983} & 2.32 & 0.779526 & 203.3890(12) & 6\\
 \hline
 \rule{0pt}{2.5ex} 1983~\cite{Mills1975}\cite{Mills1983} & 3.252760 & 0.925110 & 203.3875(16) & 8\\
 \hline
  \rule{0pt}{2.5ex} 1984~\cite{Ritter1984} & 2.32 & 0.779526 & 203.38910(74) & 4 \\
 \hline
  \rule{0pt}{2.5ex} 2014~\cite{Ishida2014} & 2.8566 & 0.8661288 & 203.3942(21) & 10 \\
 \hline
  \rule{0pt}{2.5ex} 2015~\cite{Miyazaki2015} & 0 & 0 & 203.39(19) & 934 \\
 \hline
 
\end{tabular}}
\caption{Data reported by the most recent experiments whose goal was to measure $\Delta\nu$. For each of them, we list the transition frequency $f_0$, the magnetic field intensity $B$, the hyperfine splitting $\Delta\nu$ reported and its relative error $\delta\Delta\nu/\nu$.}
\label{tab:exp}
\end{table}

Actually, to recover the value of $\Delta\nu$, experimentalists make use (see table~\ref{tab:exp} and references therein) of the truncated theoretical expression
\begin{equation}\label{gpos}
g_{\mathrm{Ps}} = g_e\left(1-\frac{5}{24}\alpha^2\right)=2.0022971(1),
\end{equation}
where the relative uncertainty of 0.05~ppm is the rounded estimate of the known magnitude of the neglected terms in Eq.~\ref{gposCombined}.

The reason of this somewhat traditional choice, that arbitrarily fixes one fundamental constant of the two, has never been stated openly, but one may guess that is because of the fact that the theoretical prediction of $g_{\mathrm{Ps}}$ looks more precise than the theoretical prediction of $\Delta\nu$. 

Obviously, this does not apply to the recent measurement of $\Delta\nu$ at $B=0$~T~\cite{Miyazaki2015} which is not affected by the $g_{\mathrm{Ps}}$ value, but for experimental reasons the directly measured $\Delta\nu$ is given with a much higher relative error with respect to the values given by the other experiments (see table~\ref{tab:exp}).

\begin{figure}
\centering
\includegraphics[width=\linewidth]{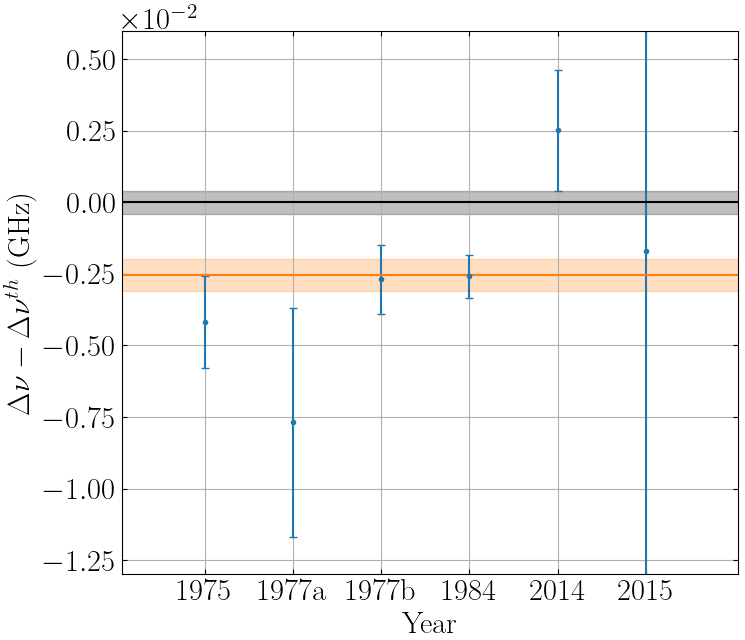}
\caption{The different experimental values reported for $\Delta\nu$, also in table~\ref{tab:exp}, ordered following their publication year. For the 2015 one, its error bar is too big to be whole contained in the figure. We also show in the figure in black the error bar of the theoretical expectation, and in red the weighted means for the six $\delta\nu$ values.}
\label{fig:expres}
\end{figure}

In Fig.~\ref{fig:expres} we show the different experimental values reported for $\Delta\nu$, also in table~\ref{tab:exp}, ordered following their publication year. For the one measured at $B=0$~T, its error bar is too big to be whole contained in the figure. We also show in the figure the error bar of the theoretical expectation.

Let's note that the weighed mean of the six $\delta\nu$ values, shown on the figure in red, is $\Delta\nu=(203.38914\pm 0.00056)$~GHz which is about 3$\sigma$ far from the theoretical prediction. This seems to indicate an important discrepancy between theory and experiment in the framework of QED.

%<<<<<<<<<<<<<<<<<<<<<<<
% Insertion about the quadratic Zeeman effect as possible explanation of the discrepancy

Various paths have been investigated in an attempt to explain this discrepancy: systematic errors,  diamagnetic effects, etc. A specific class of diamagnetic interaction terms have been shown in~\cite{feinberg} to contribute to the hyperfine energy levels of the ground state of Ps by up to 100~kHz for magnetic field $B$ of the order of 1~T. However, these terms shift by exactly the same amount $\Delta E_{Fm_F}$ all four hyperfine levels, and have no impact on the difference $f_0$~\cite{tsy}. Following~\cite{shiga}, we focus instead our attention on the interference of spin with diamagnetic interaction terms, which in second order of perturbation theory affect $f_0$. 
%<<<<<<<<<<<<<<<<<<<<<<<<<<<
% Insertion
To this end, we add to the positronium Hamiltoniam the diamagnetic interactions, which we split into scalar and tensor parts $H_{d0}$ and $H_{d2}$~\cite{shiga}
\begin{equation}\label{hfull}
H_{\rm Ps}=H_{\rm BP}+H_A+H_Z+H_{d0}+H_{d2},
\end{equation}
with
\begin{equation}
\begin{split}
H_{d0}&=\frac{e^2}{24m}\mathbf{r}^2\mathbf{B}^2,\\ 
H_{d2}&=-\frac{e^2}{16m}\left((\mathbf{r}\cdot\mathbf{B})^2-
\frac{1}{3}\mathbf{r}^2\mathbf{B}^2\right),
\end{split}
\end{equation}
where $\mathbf{r}$ is the vector joining the electron and the positron. The second-order contribution to the energy level of the $|Fm_F\rangle$ hyperfine component of the ground state of Ps, due to the scalar parts of the interaction terms, includes contributions from the excited states in the discrete and the continuous spectrum of positronium:
\begin{equation}\label{2nd-sc}
\Delta^{\rm (sc)}E_{Fm_F}=\Delta^{\rm (sc)}_{\rm disc}E_{Fm_F}+
\Delta^{\rm (sc)}_{\rm cont}E_{Fm_F}
\end{equation}
with
\begin{equation}
\begin{split}
\Delta^{\rm (sc)}_{\rm disc}E_{Fm_F}&=2\sum\limits_{n>1}
\langle 1s,Fm_F|H_{sc}|ns,Fm_F\rangle\\
&\times\frac{1}{E_{1}-E_{n}}
\langle ns,Fm_F|H_{d0}|1s,Fm_F\rangle,
\end{split}
\end{equation}%
and
\begin{equation}
\begin{split}
\Delta^{\rm (sc)}_{\rm cont}E_{Fm_F}&=2\int\limits_0^{\infty} dk\,
\langle 1s,Fm_F|H_{sc}|k,Fm_F\rangle\\
&\times\frac{1}{E_{1}-E(k)}
\langle k,Fm_F|H_{d0}|1s,Fm_F\rangle.
\end{split}
\end{equation}
where the ``scalar interaction term'' $H_{sc}=H_F+H_A$ (counterpart of the scalar diamagnetic interaction term $H_{d0}$) is the sum of the Fermi spin-spin interaction of the electron and the positron $H_F=(2\pi/3) \left(e\hbar g_e/mc\right)^2(\mathbf{S}\cdot\mathbf{I}) \delta(\mathbf{r})$ and the annihilation term $H_{A}$, $E_{n}=-mc^2\alpha^2/4n^2$ is the Coulomb energy of Ps in the $(n)$-state, while $E(k)=k^2 mc^2\alpha^2/4$ is the energy of the continuous spectrum state with wave vector $k$. 
The matrix elements contributing to $\Delta^{\rm (sc)}_{\rm disc}E_{Fm_F}$ may be evaluated in closed form and the numerical summation over $n$ yields 
\begin{equation}\label{eq:sc_disc}
\left(\frac{\Delta^{\rm (sc)}_{\rm disc}E_{Fm_F}/h}{\mathrm{Hz}}\right)\approx
\bigg(114\, F(F+1)-98\bigg)\left(\frac{B}{\mathrm{T}}\right)^2,
\end{equation}
where the magnetic field is assumed to be in units T, and numerical coefficient is adjusted to return the energy shift in Hz. 
The continuous spectrum contribution $\Delta^{\rm (sc)}_{\rm cont}E_{Fm_F}$ was evaluated numerically and proved to be
\begin{equation}\label{eq:sc_cont}
\left(\frac{\Delta^{\rm (sc)}_{\rm cont}E_{Fm_F}/h}{\mathrm{Hz}}\right)\approx 
\bigg(78\,(F(F+1)-67\bigg)\left(\frac{B}{\mathrm{T}}\right)^2.
\end{equation}
The scalar correction $\Delta^{\rm (sc)}E_{Fm_F}$ is independent of the quantum number $m_F$, shifts all hyperfine levels $F=1$, $m_F=-1,0,1$ by the same amount, and therefore does not affect $f_0$. It does contribute to the hyperfine splitting $\Delta\nu$ by an amount of the order of 383~Hz, which is, however, much below the current uncertainty of $\Delta\nu$. 

The tensor part $\Delta^{\rm {\rm (t)}}E_{F m_F}=\Delta^{\rm (t)}_{\rm disc}E_{F m_F}+\Delta^{(t)}_{\rm cont}E_{F m_F}$ of the second-order contribution to the energy level of the $\vert Fm_F\rangle$ only affects states with $J=F=1$:
\begin{equation}\label{2nd-t}
\begin{split}
\Delta^{\rm (t)}_{\rm disc}E_{1m_F}&=2\sum\limits_{n>2}
\langle 1011m_F|H_{t}|n211m_F\rangle\\
&\quad\times\frac{1}{E_{1}-E_{n}}
\langle n211m_F|H_{d2}|1011m_F\rangle,
\end{split}
\end{equation}
and similar  for the continuous spectrum contribution. Here
\begin{equation}
H_t=\frac{3}{4}\left(\frac{e\hbar g_e}{mc}\right)^2
\frac{1}{r^5}
\left(
(\mathbf{r}\cdot\mathbf{S})
(\mathbf{r}\cdot\mathbf{I})-
\frac{1}{3}\mathbf{r}^2(\mathbf{S}\cdot\mathbf{I})\right).
\end{equation}

This leads to  second order corrections the energy levels of the triplet $F=1$ ground state $\Delta^{\rm (t)}E_{1m_F}$ which depend on $m_F$ and, therefore, contributes to $f_0$:
\begin{equation}
\left(\frac{\Delta^{\rm (t)}E_{1m_F}/h}{\mathrm{Hz}}\right)\approx 0.07\times C_{1m_F,20}^{1m_F}\left(\frac{B}{\mathrm{T}}\right)^2,
\end{equation}
where $C_{1m_F,20}^{1m_F}$ is a coefficient of Clebsch-Gordan, 
$C_{10,20}^{10}=-2C_{1\pm1,20}^{1\pm1}=-1/\sqrt{10}$.
For magnetic fields of the order of 1~T, however, $\Delta^{\rm (t)}E_{1m_F}$ is extremely small and cannot be experimentally detected. This negative results should definitely close the search for explanation of the discrepancies in hyperfine spectroscopy of positronium with diamagnetism effects.

%>>>>>>>>>>>>>>>>>>>>>>>

\begin{figure}
\centering
\includegraphics[width=\linewidth]{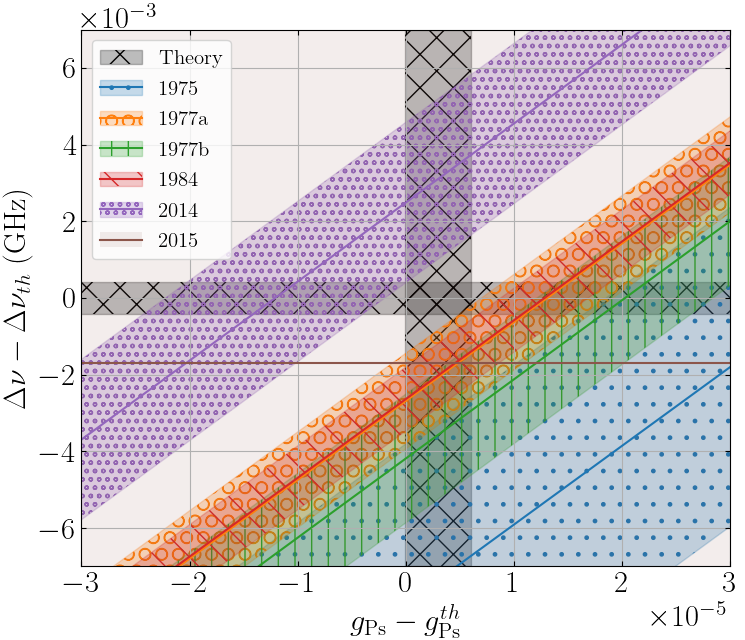}
\caption{The $\Delta\nu$ versus $g_{\mathrm{Ps}}$ $1\sigma$ accepted areas corresponding to the existing measurements of table \ref{tab:exp}. We also show the theoretical value of $\Delta\nu$ and $g_{\mathrm{Ps}}$ with their error bars.}
\label{fig:gVSdnu}
\end{figure}

Now, to go further let's unfreeze the value of $g_{\mathrm{Ps}}$ and express $\Delta\nu$ as a function of $g_{\mathrm{Ps}}$ to see, for example, if a different choice of $g_{\mathrm{Ps}}$ may solve the observed discrepancy
\begin{equation}\label{eqDeltaNuVSgpos}
\Delta\nu = f_0\left[\left(\frac{g_{\mathrm{Ps}}\mu_BB}{hf_0}\right)^2-1\right].
\end{equation}
Fig.~\ref{fig:gVSdnu} shows the $\Delta\nu$ versus $g_{\mathrm{Ps}}$ $1\sigma$ accepted area corresponding to the existing measurements of table~\ref{tab:exp}. The 2015 experiment gives an inclusion area parallel the $g_{\mathrm{Ps}}$ axis, since it does not depend on the $g_{\mathrm{Ps}}$ value and it exceeds the dimension of the figure. 

On the axis $g_{\mathrm{Ps}}=g_{\mathrm{Ps}}^{th}$ we find again the $\Delta\nu$ given in the papers, and on the axis $\Delta\nu=\Delta\nu_{th}$ we find the values of $g_{\mathrm{Ps}}$ that experimentalists would have worked out if they would have taken the choice to fix $\Delta\nu$ instead of $g_{\mathrm{Ps}}$. 

In Fig.~\ref{fig:gVSdnu} we also show the theoretical value of $\Delta\nu$ with its error bar and the theoretical value of $g_{\mathrm{Ps}}$. As for its error bar, we have taken $1\times 10^{-7}$ for the left side to take into account the difference between formula~\ref{gposSeb} and \ref{gposLew}. For the right side the difference between the two formulas is even bigger because following~\cite{Ishida2014} one has to take into account that positronium can be formed and annihilate at center of mass energies as big as 3~eV. This corresponds up to a shift of $6\times 10^{-6}$ of the $g_{\mathrm{Ps}}$ expected value that we show in the figure \ref{fig:gVSdnu}, as well.
It is important to stress also that, always in~\cite{Ishida2014}, it is  stated that previous experiments not having considered temperature effects have underestimated their error bars.

\begin{figure}
\centering
\includegraphics[width=\linewidth]{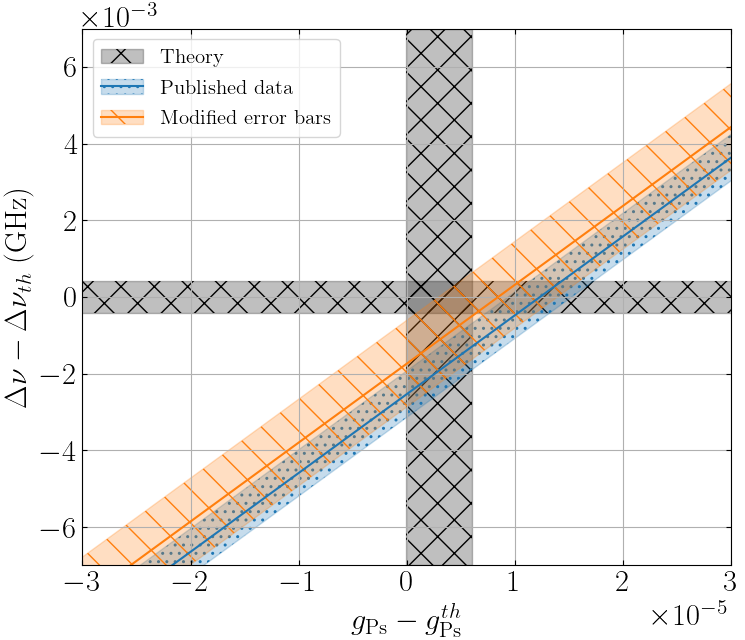}
\caption{The weighted mean for $\Delta\nu$ versus $g_{\mathrm{Ps}}$ $1\sigma$ accepted area corresponding to the existing measurements of table \ref{tab:exp} with error bars increased as suggested in ref. \cite{Ishida2014}. We also show the non corrected weighted mean and the theoretical value of $\Delta\nu$ and $g_{\mathrm{Ps}}$ with their error bars.}
\label{fig:meangVSdnu}
\end{figure}

We see that no region can be delimited by the superposition of the accepted areas of each measurement. The reason is that the variation of $\Delta\nu$ with respect to the variation of $g_{\mathrm{Ps}}$ is linear around $(g_{\mathrm{Ps}}^{th}, \Delta\nu_{th})$ 
\begin{equation}\label{DerDeltaNuVSgpos}
\frac{d(\Delta\nu)}{dg_{\mathrm{Ps}}} = \left(\frac{2g_{\mathrm{Ps}}^{th}\mu_B^2B^2}{h^2f_0}\right).
\end{equation}
All the experiments except one have been performed at about 1~T field and therefore a similar few GHz $f_0$ has been measured. Their areas of inclusion are therefore parallel at the lowest order. This means that the $\Delta\nu$ value given by experiments is as arbitrary as the choice of to fix $g_{\mathrm{Ps}}$. For example, if one fixes $g_{\mathrm{Ps}}$ at a value about $1\times10^{-5}$ higher than $g_{\mathrm{Ps}}^{th}$, i.e. one doubles the $\alpha^2$ correction to $g_e$, one gets a $\Delta\nu$ average value in agreement with the theoretical prediction (see Fig.~\ref{fig:meangVSdnu}).

For the sake of the argument, let's do what reference~\cite{Ishida2014} suggests. We increase quadratically the error bar of the oldest experiment by 10 ppm, calculate the weighted mean as function of $g_{\mathrm{Ps}}$. The result is shown in Fig.~\ref{fig:meangVSdnu}.

We see that the weighted mean thus obtained is compatible with the theoretical expectations of both $\Delta\nu$ and $g_{\mathrm{Ps}}$ opening a possibility to explain the observed discrepancy between theory and experiment by the temperature effects.

Clearly, the only way to give both fundamental constants in an unquestionable way is to perform measurement at different magnetic field values and to combine the results obtained. To illustrate our statement, let's recall that a measurement at $B=0$~T would corresponds to an inclusion region in Fig.~\ref{fig:gVSdnu} parallel to the $\Delta\nu$ axis while a measurement at a field of 40~T would correspond to an inclusion area  almost parallel to the $g_{\mathrm{Ps}}$ axis since at very high field $f_0\approx g_{\mathrm{Ps}}\mu_BB/h$, therefore independent on $\Delta\nu$.

Can we give anyway a value for both constant using the existing results? The answer is yes but eventually their precision will be disappointing. 

\begin{figure}
\centering
\includegraphics[width=\linewidth]{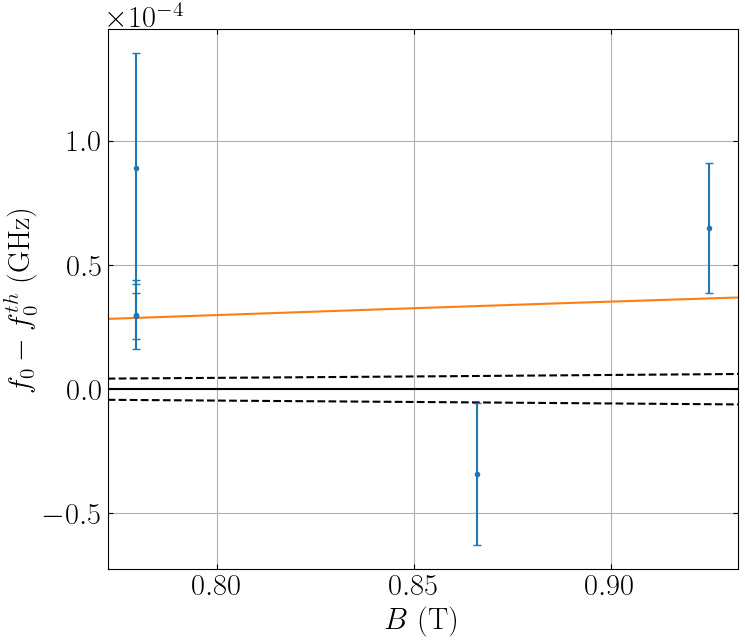}
\caption{$f_0$ versus $B$(T) corresponding to table~\ref{tab:exp}, theoretical curve in black and curve obtained by best fitting in red.}
\label{fig:f0SB}
\end{figure}

In Fig.~\ref{fig:f0SB} we show $f_0$ versus $B$(T) corresponding to table~\ref{tab:exp}. The error bars associated to each point is obtained by assuming the same relative error on $\Delta\nu$ for $f_0$. We have excluded the point at $B=0$~T, again because of its big error bar. To guide the eyes we also show the expected theoretical curve \emph{i.e.} the $f_0$ vs $B$ curve obtained injecting $g_{\mathrm{Ps}}^{th}$ and $\Delta\nu_{th}$ in Eq.~\ref{Breit}.

We can therefore fit the experimental points with Eq.~\ref{Breit}. We obtain $g_{\mathrm{Ps}}=(2.0021\pm 0.0018)$ and $\Delta\nu=(203.34\pm 0.38)$~GHz. The corresponding curve is shown in red in Fig.~\ref{fig:f0SB}. Precision is very low, we are fitting few points with a two free parameters function. %We obtain $g_{\mathrm{Ps}}=(2.0008\pm 0.0047)$ et $\Delta\nu=(203.07\pm 0.97)$~GHz.

Finally, since a measurement of $\Delta\nu$ at $B=0$~T has been published, as a matter of fact, we can disentangle $\Delta\nu$ and $g_{\mathrm{Ps}}$. We can inject the experimental value of $\Delta\nu$ in Eq.~\ref{Breit} and extract a value of $g_{\mathrm{Ps}}$ for any measurement at $B \neq 0$. All the experiments give a $g_{\mathrm{Ps}} = 2.0023\pm 0.0027$ at $3\sigma$ and therefore the average value is $g_{mean}=2.0023\pm 0.0012$ at $3\sigma$. The error $\delta g_{\mathrm{Ps}}$ is limited by the error of the measurement of $\Delta\nu$, $\delta \Delta\nu$, at $B=0$~T, which means that any new more precise measurement of $\Delta\nu$ at $B=0$~T will also increase our knowledge of the experimental value of $g_{\mathrm{Ps}}$. 
Actually, one can write 
\begin{equation}\label{dSgpos}
\frac{\delta g_{\mathrm{Ps}}}{g_{\mathrm{Ps}}} = \frac{\delta \Delta\nu}{\Delta\nu}\left(\frac{1}{2(1+f_0/\Delta\nu)}\right),
\end{equation}
with the factor in parenthesis going from 1/2 to 0 when $f_0/\Delta\nu$ goes from 0 to $\infty$.

All the values are in agreement with the expected theoretical value $g_{th}=2.0023$ within the experimental error. This value is much less precise than the reported values of $\Delta\nu$ but it is unquestionable and, one could say, it is the only available. 
 
In conclusion, in this letter we have shown why the value of the g factor of the positronium is important. Taking this better into consideration may provide a solution to the reported discrepancy between QED theory and experiment concerning the hyperfine splitting of the fundamental level of the positronium.  We also give the only experimental value that existing experiment can provide $g_{\mathrm{Ps}}=2.0023\pm 0.0012$ at $3\sigma$. The precision is certainly disappointing concerning QED tests. 

Our work suggests that the current status of the theory and experiments are quite incomplete. As for experiments, clearly, we need more precise experiments at $B=0$~T and experiments at high $B$ like 10~T or 40~T and even more, using therefore latest developments in high magnetic field technology \cite{HIMAFUN}. 
%<<<<<<<<<<<<<<<<<<<<
% Insertion
The scalar diamagnetic correction to the triplet energy levels 
$\Delta^{\rm (sc)}E_{F m_F}$ Eq.~\ref{eq:sc_disc} and \ref{eq:sc_cont} increases quadratically with $B$, and for such fields will give an experimentally observable contribution to $\Delta\nu$ and will need to be taken into account. 
On the other hand, above $B\sim 3$~T the relative contribution of $\Delta^{\rm (sc)}E_{F m_F}$ to $f_0$ (see Eq.~\ref{Breit}) increases only linearly with $B$ and can be neglected even at higher magnetic fields.
%>>>>>>>>>>>>>>>>>>>>

THz sources exist~\cite{Miyazaki2015}, optical transitions of a Rubidium gas has been observed up to about 60~T~\cite{RUHMA} in the framework of magnetic field metrology and the formation of positronium in a rubidium target has been largely studied (see \emph{e.g.}~\cite{Pandey2016} and references within). At least in principle, it looks like there is no objection for such a novel experiment that could provide a first precise measurement of both the hyperfine splitting and the g-factor of positronium. As far as we are concerned, since we are willing to go further, we will study in details the feasibility of such an experiment in the near future.

At the end let us briefly discuss the theory-experiment discrepancy in the fine structure of the $n = 2$ level of positronium reported in~\cite{Gurung2020}. The analysis of the experimental data in this case, which is beyond the scope of the present paper, is based on an expression that -- unlike Eq.~\ref{Breit} -- involves several state-dependent g-factors of the 2S- and 2P-states~\cite{Lewis1973}. Their values, however, are not clearly referenced that, in our understanding, is an important piece of information lacking. 
We only note here that the diamagnetic effects in $n=2$ states are more pronounced because they contribute to the $2s-2p$ energy gap in first order of perturbation theory.  Indeed, the correction to the Coulomb energy level of the $(n,l)$ state of positronium due to $H_{d0}$ is
\begin{equation}
\begin{split}
\Delta^{d0}E_{nl}&=\langle nl|H_{d0}|nl\rangle\\
&=K \left(5n^4+n^2(1-3l(l+1))\right)\left(\frac{B}{\mathrm{T}}\right)^2
\end{split}
\end{equation}
where $K=2.48$~kHz. This contributes by approximately 30 kHz $(B/T)^2$ to the $2s-2p$ energy difference, and gives a correction of the order of  $10^{-6}$ to the parameter $a$ in the fit to the experimental data of~\cite{Gurung2020} (see the caption to Fig.~4 therein), which is too small to even partially explain the observed discrepancy.

\section*{Statements and Declarations}
\paragraph{Acknowledgments} One of the authors (D.B.) is acknowledging the partial support from Grant No. KP-06-N58/5 of the Bulgarian Fund for scientific research. Another author (J.A.) is acknowledging support from the ``Mission pour les Initiatives Transverses et Interdisciplinaires'' of the CNRS.

\paragraph{Author contributions} All the authors were involved in the preparation of the manuscript. All the authors have read and approved the final manuscript.

\paragraph{Data Availability Statement} The datasets generated during and/or analyzed during the current study are available from the corresponding author on reasonable request.

%\begin{figure}
%% Use the relevant command for your figure-insertion program
%% to insert the figure file.
%% For example, with the option graphics use
%\resizebox{0.75\textwidth}{!}{%
%  \includegraphics{leer.eps}
%}
%% If not, use
%%\vspace{5cm}       % Give the correct figure height in cm
%\end{figure}
%
%% For two-column wide figures use
%\begin{figure*}
%% Use the relevant command for your figure-insertion program
%% to insert the figure file. See example above.
%% If not, use
%\vspace*{5cm}       % Give the correct figure height in cm
%\end{figure*}
%\begin{table}
%\begin{tabular}{lll}
%\hline\noalign{\smallskip}
%\end{tabular}
%% Or use
%\vspace*{5cm}  % with the correct table height
%\end{table}
%
% BibTeX users please use
% \bibliographystyle{}
% \bibliography{}

\end{document}